\documentclass[runningheads,a4paper]{llncs}
\usepackage{cite}
\usepackage{paralist}
\usepackage{todonotes}
\usepackage[hyphens]{url}

\begin{document}
\title{Integrating Privacy-Enhancing Technologies into the Internet Infrastructure}
\author{David Harborth\inst{1} \and Dominik Herrmann\inst{2} \and Stefan K{\"{o}}psell\inst{3} \and Sebastian Pape\inst{1} \and Christian Roth\inst{4} \and Hannes Federrath\inst{2} \and Dogan Kesdogan\inst{4} \and Kai Rannenberg\inst{1}}
\institute{Goethe-University Frankfurt am Main \and University of Hamburg \and TU Dresden  \and University of Regensburg}
\maketitle

\begin{abstract}
The AN.ON-Next project aims to integrate privacy-enhancing technologies into the internet's infrastructure and establish them in the consumer mass market.\\ 
The technologies in focus include a basis protection at internet service provider level, an improved overlay network-based protection and a concept for privacy protection in the emerging 5G mobile network.
A crucial success factor will be the viable adjustment and development of standards, business models and pricing strategies for those new technologies. 
\end{abstract}

\section{Introduction}
\label{intro}
Despite an increasing public perception of the matter of data protection, nowadays anonymization services like Tor and JonDonym have not yet achieved wide everyday and mass appeal. Although a user base of tens of thousands (JonDonym) to several hundreds of thousands (Tor) users is a decent result, their share is vanishing small compared to the total number of internet users~\cite{torUsers}.

As a result, most internet users today leave extensive digital traces that can be used to build detailed personal profiles by internet service providers (ISP) or third parties without the users' knowledge and possibility of intervention. This threat to the right of informational self-determination gains in importance through the pervasion of everyday's life by the internet.
In particular, the increasing use of portable devices
leads to the possibility of an even more detailed profiling and thus allows a deeper intrusion in the users' privacy.

Otherwise, one important reason for the low prevalence of privacy-enhancing technologies (PETs) is the lack of usability. Privacy played no significant role in the design of today's internet's infrastructure. Thus, actual anonymization services are organized as separate overlay networks. End users typically need to install additional components on their system. This results in a massive effort to use those programs which in turn leads to an overburdened user. Additionally, the usability of Tor and JonDonym is limited to stationary personal computers and comparable software for mobile devices is not available for the consumer mass market yet. On the other hand, many PETs cause a high overhead and cannot be activated by default (e.g. by the ISP). The loss of comfort by far outweighs the benefits of privacy and is thus unacceptable for many users.

The underlying assumption that guides the project is that PETs are only able to reach the mass market when they are standardized and usable without any action of the user (``zero-effort'') and work so efficiently that they do not cause any noticeable limitation on the quality of service (especially regarding latency and bandwidth). 
This is particularly important considering the privacy attitudes and behaviors of regular users \cite{Berendt2005, Acquisti2005}. To reach this goal,  PETs need to be firmly integrated in the internet's infrastructure.

Therefore, the AN.ON-Next\footnote{\url{https://www.anon-next.de/}} project's vision is to integrate PETs in the internet's infrastructure. By pilot projects with industry representatives the concepts will be tested, optimized and shown that efficient data protection is possible and can be the basis for new business models. Up to now there are no such solutions, partly because of the lack of suitable cooperation between research and industry. This led repeatedly to missed opportunities regarding a possible incorporation of strong privacy protecting technologies in new telecommunications' and internet's standards. In order to establish PETs in the mass market, therefore corresponding business models also have to be regarded.

The project aims at working on three overarching goals. First, design more efficient PETs. Second, allocate the anonymization service into the internet's infrastructure and create a transparent anonymization experience for end-users. Third, improve the traceability of protection level achieved for each participant.

The remainder of this paper is organized as follows: Sect.~\ref{relwork} provides an overview of related work. The goals for the three tackled technical areas of ISP-based, network overlay-based, and 5G network anonymization techniques are presented in Sect.~\ref{isp} to Sect.~\ref{5g}. The development of business models is discussed in Sect.~\ref{businessmodels}. We conclude the sketch of our project idea in Sect.~\ref{conclusion}.
 

\section{Related Work}
\label{relwork}

The planned work on ISP-sided anonymization will be built on existing work by the project partners \cite{stefanIpV6,dominikIpV6}. This work will be expanded, optimized and tested based on a prototype in order to develop usable solutions.

For network overlay-based anonymization technologies exist already practically usable solutions, namely Tor and JonDonym (former JAP). The main problems of these existing services are massive performance limitations\footnote{cf. \url{https://www.anonym-surfen.de/status/}; \url{https://metrics.torproject.org/}} \cite{torSlow09,distPerfMeasureTor12}
and a lack of compatibility with existing application software. How to improve the performance is already discussed in numerous papers~\cite{gMixESORICS,encDNSesorcis14,reardon09torDtls,layer3MixingImpl09,pctcp13}.


The majority of research considers only the design of anonymization protocols for the application layer. Relevant preliminary work of the project partners are methods for protecting the location of mobile subscribers \cite{Fede99}, the gMix framework, with which anonymization protocols can be evaluated realistically in a short time \cite{gMixESORICS,gMixWeb}, proposals for lightweight anonymization protocols \cite{encDNSesorcis14}, as well as the assessment of the effectiveness of different attack techniques \cite{vinhCTAnalysis11} in order to make effective protection mechanisms and to explore their limits.

There is a rich strand of literature that deals with the development and adaption of business models to the ever changing environment companies have to deal with \cite{Spieth2014}. The groundwork in that area was laid by Staehler \cite{Staehler2002} who proposed an approach where a business model consists of three elements: value proposition, value chain (value creation architecture) and revenue model. Other fundamental work on business models was done by Wirtz and Osterwalder and Pigneur \cite{Osterwalder2010, Wirtz2001, Wirtz2013}. Furthermore, the application of business model approaches to anonymization and identity management services is considered \cite{kr10, kr11, privacyEnablers11, veseli2015}.


\section{ISP-based Anonymization}
\label{isp}
One objective in AN.ON-Next is to study light-weight mechanisms that increase the baseline protection for ordinary users. With existing approaches like Tor or I2P users have to install and run a client software on their own. In contrast, we are interested in unobtrusive techniques that minimize effort for the user. We have observed that many users are willing to accept the fact that their ISP can analyze their surfing behavior, while they object to tracking performed by ad networks and profiling services.

Existing protection techniques, such as deleting cookies and preventing browser fingerprinting, are ineffective, if the traffic of a user is coming from the same IP address over long periods of time. In this case third parties could link a user's activities solely based on the IP address. However, obtaining a new IP address from the ISP is a cumbersome task at the moment. Typically, one has to manually force the broadband router to perform a reconnect, which terminates all active connections. The situation will worsen in the future, if ISPs decide to assign a long-lasting IPv6 prefix to residential customers.

In principle, ISPs could offer basic privacy protection with little cost. To this end, ISPs would only assign very short-lived IP addresses (or IPv6 prefixes) to their customers. This measure would complement defenses that are already implemented in major browsers by ensuring that they are not bypassed with IP-based tracking efforts. However, network protocols used during dial-up (e.\,g., DHCP and PPPoE) have not been designed with short-lived addresses in mind.

Additionally we will investigate to which extent it is feasible to have short-lived IP addresses not only on a per device base but on a per connection base, i.e. utilizing different source IP addresses per packet flow.

Therefore, we will look into various design alternatives to deploy short-lived IP addresses and study their feasibility. For instance, ISPs could employ carrier-grade network address translation in order to rewrite the traffic of a customer on their own, resulting in zero effort for the user. Alternatively, ISPs could assign multiple IPv6 prefixes to the customer's broadband router at a given point in time \cite{dominikIpV6,stefanIpV6}. In this case, the customer would still have some control about the anonymization process, because now it is the broadband router that decides which IPv6 prefix should be used for a particular outgoing connection.

\section{Network Overlay-based Anonymization}
\label{networkoverlay}
The main problems of the existing network overlay-based anonymization services are the weak performance, missing protection against strong attackers, and the high effort for users to install and use such systems. 

The developed anonymization service will be based on the concept of cascades instead of free sequences. A cascade is a fixed sequence of connected intermediate stations (mixes). The user will only be able to decide which cascade he wants to use. The use of cascades instead of free sequences aims to avoid some disadvantages of Tor associated with the selection of Tor nodes in a route.

In addition, transparency about the hosts of the mixes is not always given. Thus, the new anonymous protocol will be designed and integrated into a test cascade of the anonymization service JonDonym. The protocol should provide reliable protection against much stronger attackers than the ISP-based solution and the usability and compatibility should significantly improve compared to Tor and the current JonDonym service. A high level of transparency will be achieved by providing reliable information on the operator of each mix and other relevant data to the user. Thus, the user can decide if it is the appropriate mix cascade.

The basic idea of ​​the proposed protocol is a paradigm shift compared to current anonymization services. The solutions in focus strive to receive user data at the IP layer (instead of the application layer) over a virtual private network (VPN) connection. Therefore one interface to the new anonymization service will be a user operated JonDonym-client acting as a VPN server on a computer in his own home, for example on a wireless router. The JonDonym-client passes the communication through the (redesigned) JonDonym mix cascade for the purpose of anonymization and for example for recursive encryption and decryption. Another approach will be to run the JonDonym-client directly on the (mobile) device of the end user and let the JonDonym-client additionally act as a VPN-service. Being a VPN-service implies that the JonDonym-client will be responsible for handling the IP traffic of the mobile device which the JonDonym-client will tunnel utilizing the anonymization service (instead of a usual VPN (e.g. IPSec based) as an ordinary VPN service would do). 

The advantage is that the concept produces compatibility with all VPN enabled devices including the installed software. The user only has to add the JonDonym-client as a VPN server to his terminal, which is supported by all major smartphone platforms already. This will reduce the configuration effort to a minimum and creates compatibility with devices and applications which was not achieved by previous solutions.

\section{Anonymization Techniques for the 5G Network}
\label{5g}
Mobile networks experienced an exponential increase of mobile data traffic and of the number of connected devices over the last decade~\cite{cisco-data}. However, this explosion together with the high demand for extremely low latency real-time applications, e.g., the so-called tactile internet, video streaming, and vehicular ad hoc networks (VANETs~\cite{DBLP:journals/telsys/ZeadallyHCIH12}), impose new challenges on the current mobile networks. For instance, in VANETs, each vehicle periodically sends, receives, and broadcasts information to the vehicular network in order to increase traffic safety. The communication between the vehicles and the network as well as among the vehicles themselves rely on very accurate and up-to-date information about the surrounding environment. This in turn requires the underlying network architecture and communication protocols to provide robust connectivity and ensure fast delivery of information to all the vehicles. The next generation of mobile telecommunication, namely 5G, is therefore desirable to fulfill these requirements.

From the technical point of view, small cells are crucial in 5G networks in order to address the huge amount of data capacity. Concurrently, the deployment of small cells allows 5G networks and hence malicious attackers to localize mobile devices easier and more precisely, which renders 5G more vulnerable to location privacy threats. It is therefore pivotal to revisit the problem of location privacy in the 5G environment under the consideration of stronger adversary models.

From the architectural point of view, the conventional centralized cloud-based architectures for mobile networks may no longer be suitable to provide the 1 millisecond round trip delay that is typically a crucial requirement for many real-world scenarios such as VANETs. This challenge motivates the use of various local clouds in the design of 5G architectures. Thereby, mobile devices are strongly coupled with their local computing resources, i.e., the clouds, which allows users' location information to be distributed and replicated in the cloud databases. Due to this massive amount of data and redundancies, effectively managing location privacy in such architectures is a non-trivial task.

In this project, we are aiming to address two challenging privacy problems in 5G networks, namely location privacy and privacy management.

For location privacy, we are looking for different anonymization techniques ranging from lightweight anonymity protections, e.g, frequently changing the pseudonyms of mobile devices, to more advanced privacy protection techniques against stronger attacker models, e.g., mix-zones~\cite{DBLP:journals/pervasive/BeresfordS03,DBLP:conf/pet/FreudigerSH09}.
It is worth noting that there is a conflict between the level of location privacy protection and the optimization of service quality. In particular, service providers often require users to provide more personal data such as their birthdays or their current locations in order to support the users better. At the same time, disclosing too much information puts the users at more potential privacy risks. We therefore take this trade-off into account while looking for good location privacy protections.

We tackle the problem of privacy management in 5G from two directions. The first approach is to investigate mechanisms to specify privacy policies that prevent unauthorized access to raw location data. One possible solution could be to extend the state-of-the art techniques, e.g., EPA and EPAL~\cite{IBM-EPA,DBLP:conf/esorics/BackesPS03} to the context of 5G. In the second approach, we exploit different transparency enhancing techniques (see ~\cite{DBLP:conf/stast/JanicWV13} for an overview) that provide users with information on how their data is being processed, stored, disclosed, and so forth. These techniques enable users to protect their own privacy by choosing appropriate actions.

Additionally due to the high demands regarding latency, bandwidth, number of users per cell etc. many of the existing PETs cannot be simply adopted to the 5G setting. Anonymous communication via distributed solutions like Tor or JonDonym with the aim of end-to-end latency not higher than 1ms implies that the anonymization servers (mixes) have to be physically located within a 150km radius from the mobile device (because of the speed of light).

\section{Business Models for Privacy-Enhancing Technologies in the Internet Infrastructure}
\label{businessmodels}
The success of the developed technologies depends heavily on the wide distribution in the consumer mass market. This can only be achieved by creating a suitable business model that evolves around the technologies and regards the interests of all relevant stakeholders. Therefore, the business model generation is one important step in this research.

Based on related work, it will be investigated whether there are ways to adapt existing business models. The prevalent goal is to create profitable and sustainable ways for implementing and operating the developed anonymization services in order to incentivize ISPs to engage in this business. This is the condition for ensuring a rapid and wide spread of the technologies.

In addition, it will be ensured that the achieved technical solutions are feasible from an economical perspective by developing business models together with the PETs in an iterative way. The focus will be on the design of business models that enable the ISP-sided anonymity for all customers of the ISPs. It is investigated whether there are ways of cutting costs in the operation of the anonymization service infrastructure (an example could be the direct operation of Mix servers on the Internet backbone).

In a next steps, various tariff plans are examined to determine to what extent they are suitable for refinancing the concepts and to estimate which stakeholders must carry economic risks in certain scenarios. Tariff models have several different properties with different characteristics that all must be considered for the project. For example: 
\begin{compactitem}
\item Billing models: one-off payments, flat fees, consumption-based pay, financing by advertisement or consumption-related charges
\item Quality levels of service: differentiation in terms of speed or different privacy protection levels
\end{compactitem}
These analyses are carried out on a target scenario compared to the status quo for all different technologies. Those results are also discussed iteratively with the developers of the technologies to identify optimization potential.

The customer acceptance and understanding of tariff models determines crucially how successful a business model will be. Therefore possible tariff models are investigated with regard to usability, i.e. whether the end user understands the tariff and whether it is possible for the end user to choose an appropriate tariff in line with his needs. The main research problem in this area is how end users perceive the various service features with regard to the associated prices of the services.

\section{Conclusion}
\label{conclusion}
We sketched three different areas of PETs along with proposals how to improve their usability and/or performance. Additionally, we described ideas how to integrate business models into technological research when integrating the PETs in the internet infrastructure. We assume, that all fields of activity (usability and performance improvement, business models) are needed to achieve our goal to bring PETs in the consumer mass market for internet access.

\section{Acknowledgements}
\label{acknowledgements}
The project is funded by the German Federal Ministry of Education and Research (BMBF) via the program ``self-determined and secure in the digital world''.

\bibliography{quellen}

\begin{thebibliography}{10}

\bibitem{Acquisti2005}
A.~Acquisti and J.~Grossklags.
\newblock {Privacy and rationality in individual decision making}.
\newblock {\em IEEE Security and Privacy Magazine}, January/February(1):24--30,
  Jan 2005.

\bibitem{pctcp13}
Mashael AlSabah and Ian Goldberg.
\newblock {PCTCP: Per-Circuit TCP-over-IPsec Transport for Anonymous
  Communication Overlay Networks}.
\newblock In Ahmad-Reza Sadeghi, Virgil~D. Gligor, and Moti Yung, editors, {\em
  {CCS'13}}, pages 349--360. ACM, 2013.

\bibitem{DBLP:conf/esorics/BackesPS03}
Michael Backes, Birgit Pfitzmann, and Matthias Schunter.
\newblock A toolkit for managing enterprise privacy policies.
\newblock In {\em Computer Security - {ESORICS} 2003, 8th European Symposium on
  Research in Computer Security, Proceedings}, pages 162--180, 2003.

\bibitem{Berendt2005}
Bettina Berendt, Oliver Guenther, and Sarah Spiekermann.
\newblock {Privacy in e-commerce}.
\newblock {\em Communications of the ACM}, 48(4):101--106, Apr 2005.

\bibitem{DBLP:journals/pervasive/BeresfordS03}
Alastair~R. Beresford and Frank Stajano.
\newblock Location privacy in pervasive computing.
\newblock {\em {IEEE} Pervasive Computing}, 2(1):46--55, 2003.

\bibitem{cisco-data}
{Cisco visual networking index: Global mobile data traffic forecast update,
  2015-2020}.
\newblock
  \url{http://www.cisco.com/c/en/us/solutions/collateral/service-provider/visual-networking-index-vni/mobile-white-paper-c11-520862.pdf}.

\bibitem{torSlow09}
Roger Dingledine and Steven~J. Murdoch.
\newblock {Performance Improvements On Tor Or, Why Tor Is Slow And What We're
  Going To Do About It}.
\newblock Technical report, Tor Project, March 2009.

\bibitem{Fede99}
Hannes Federrath.
\newblock {\em Sicherheit mobiler {K}ommunikation}.
\newblock DuD Fachbeitr{\"a}ge. Vieweg, Wiesbaden, 1999.

\bibitem{stefanIpV6}
Florent Fourcot, Laurent Toutain, Stefan K{\"{o}}psell, Fr{\'{e}}d{\'{e}}ric
  Cuppens, and Nora Cuppens{-}Boulahia.
\newblock Ipv6 address obfuscation by intermediate middlebox in coordination
  with connected devices.
\newblock volume 8115 of {\em LNCS}, pages 148--160, 2013.

\bibitem{DBLP:conf/pet/FreudigerSH09}
Julien Freudiger, Reza Shokri, and Jean{-}Pierre Hubaux.
\newblock On the optimal placement of mix zones.
\newblock In {\em Privacy Enhancing Technologies, 9th International Symposium,
  {PETS} 2009, Seattle, WA, USA, August 5-7. Proceedings}, pages 216--234,
  2009.

\bibitem{gMixESORICS}
Karl-Peter Fuchs, Dominik Herrmann, and Hannes Federrath.
\newblock {Introducing the gMix Open Source Framework for Mix Implementations}.
\newblock In Sara Foresti, Moti Yung, and Fabio Martinelli, editors, {\em
  {{ESORICS}'12}}, volume 7459 of {\em LNCS}, pages 487--504. Springer, 2012.

\bibitem{dominikIpV6}
Dominik Herrmann, Christine Arndt, and Hannes Federrath.
\newblock Ipv6 prefix alteration: An opportunity to improve online privacy.
\newblock {\em CoRR}, abs/1211.4704, 2012.

\bibitem{encDNSesorcis14}
Dominik Herrmann, Karl-Peter Fuchs, Jens Lindemann, and Hannes Federrath.
\newblock {EncDNS: A Lightweight Privacy-Preserving Name Resolution Service}.
\newblock In {\em {{ESORICS}'14}}, volume 8712 of {\em LNCS}, pages 37--55,
  2014.

\bibitem{IBM-EPA}
{{E}nterprise {P}rivacy {A}rchitecture: {S}ecuring returns on e-business}.
\newblock
  \url{https://www-935.ibm.com/services/ph/bcs/pdf/g510-1913-enterprise-privacy-architecture.pdf}.

\bibitem{DBLP:conf/stast/JanicWV13}
Milena Janic, Jan~Pieter Wijbenga, and Thijs Veugen.
\newblock Transparency enhancing tools (tets): An overview.
\newblock In {\em Third Workshop on Socio-Technical Aspects in Security and
  Trust, {STAST} 2013, New Orleans, LA, USA, June 29, 2013}, pages 18--25.
  {IEEE} Computer Society, 2013.

\bibitem{layer3MixingImpl09}
Csaba Kir{\'a}ly and Renato~Lo Cigno.
\newblock {IPsec-Based Anonymous Networking: A Working Implementation}.
\newblock In {\em {{ICC}'09}}, pages 1--5. IEEE, 2009.

\bibitem{kr10}
S.~Koschinat, G.~Bal, M.~Hegen, and K.~Rannenberg.
\newblock {H6.1.2 - Towards an Economic Valuation of Identity Management
  Enablers, Public Deliverable of EU Project PrimeLife}.
\newblock \url{http://primelife.ercim.eu/results/documents/129-612h}, 2010.

\bibitem{kr11}
S.~Koschinat, G.~Bal, K.~Rannenberg, and M.~Hegen.
\newblock {D6.1.2 - Economic Valuation of Identity Management Enablers, Public
  Deliverable of EU Project PrimeLife}.
\newblock \url{http://primelife.ercim.eu/results/documents/151-612d}, 2011.

\bibitem{privacyEnablers11}
Sascha Koschinat, G{\"{o}}khan Bal, Christian Weber, and Kai Rannenberg.
\newblock Privacy by sustainable identity management enablers.
\newblock In {\em Privacy and Identity Management for Life}, pages 431--452.
  2011.

\bibitem{torUsers}
{Mark Graham and Stefano De Sabbata}.
\newblock {Information Geographies at the Oxford Internet Institute -- The
  anonymous Internet}.
\newblock \url{http://geography.oii.ox.ac.uk/?page=tor}, 2015.

\bibitem{distPerfMeasureTor12}
Sebastian M{\"u}ller, Franziska Brecht, Benjamin Fabian, Steffen Kunz, and
  Dominik Kunze.
\newblock {Distributed Performance Measurement and Usability Assessment of the
  Tor Anonymization Network}.
\newblock {\em Future Internet}, 4(2):488--513, 2012.

\bibitem{Osterwalder2010}
Alexander Osterwalder and Yves Pigneur.
\newblock {\em Business Model Generation - A Handbook for Visionaries, Game
  Changers, and Challengers}.
\newblock Wiley, New York, 2010.

\bibitem{vinhCTAnalysis11}
Dang~Vinh Pham, Joss Wright, and Dogan Kesdogan.
\newblock A practical complexity-theoretic analysis of mix systems.
\newblock volume 6879 of {\em LNCS}, pages 508--527, 2011.

\bibitem{reardon09torDtls}
Joel Reardon and Ian Goldberg.
\newblock {Improving Tor using a TCP-over-DTLS Tunnel}.
\newblock In {\em {{USENIX Sec}'09}}, pages 119--134. USENIX Association, 2009.

\bibitem{Spieth2014}
Patrick Spieth, Dirk Schneckenberg, and Joan~E Ricart.
\newblock {Business model innovation â state of the art and future
  challenges for the field}.
\newblock {\em R{\&}D}, 44(3):237--247, 2014.

\bibitem{Staehler2002}
Patrick St\"{a}hler.
\newblock {\em Geschäftsmodelle in der digitalen Ökonomie - Merkmale,
  Strategien und Auswirkungen}.
\newblock Amazon.de, Lohmar, K\"{o}ln, 2. aufl. edition, 2002.

\bibitem{gMixWeb}
{The gMix Project}.
\newblock {gMix: A Generic Open Source Framework for Mixes}.
\newblock \url{https://svs.informatik.uni-hamburg.de/gmix/}, 2015.

\bibitem{veseli2015}
F.~Veseli and W.~Tesfay.
\newblock {Privacy-ABC Technologies, Personal Data Ecosystem, and Business
  Models -- A feasibility study report. Technischer Bericht, ABC4Trust
  Project}, 2015.

\bibitem{Wirtz2001}
Bernd~W. Wirtz.
\newblock {\em Electronic Business}.
\newblock Springer-Verlag, Berlin Heidelberg New York, 2. aufl. edition, 2001.

\bibitem{Wirtz2013}
Bernd~W. Wirtz.
\newblock {\em Business Model Management - Design - Instrumente -
  Erfolgsfaktoren von Geschäftsmodellen}.
\newblock Gabler Verlag, Wiesbaden, 3. aufl. edition, 2013.

\bibitem{DBLP:journals/telsys/ZeadallyHCIH12}
Sherali Zeadally, Ray Hunt, Yuh{-}Shyan Chen, Angela Irwin, and Aamir Hassan.
\newblock Vehicular ad hoc networks {(VANETS):} status, results, and
  challenges.
\newblock {\em Telecommunication Systems}, 50(4):217--241, 2012.

\end{thebibliography}
\bibliographystyle{plain}
\end{document}